\documentclass[prb,showpacs,twocolumn]{revtex4}
\usepackage{graphicx,color}
\usepackage{amssymb}
\usepackage{rotating}

\newcommand{\half}{\frac{1}{2}}
\newcommand{\nn}{\nonumber}
\newcommand{\eqref}[1]{Eq.~(\ref{#1})}
\newcommand{\ket}[1]{\left|{#1}\right\rangle}
\newcommand{\bra}[1]{\left\langle{#1}\right|}

\begin{document}

\title{Geometric quantum gates with superconducting qubits}
\author{I. Kamleitner$^1$, P. Solinas$^2$, C. M\"uller$^{1,3}$, A. Shnirman$^{1,3}$, and M. M\"ott\"onen$^{2,4}$}
\affiliation{${}^1$Institut f\"ur Theory der Kondensierten Materie, Karlsruher Institut f\"ur Technologie, 76128 Karlsruhe, Germany}
\affiliation{${}^2$Department of Applied Physics/COMP, Aalto University, P.O.~Box 14100, FI-00076 AALTO, Finland}
\affiliation{${}^3$DFG-Center for Functional Nanostructures (CFN), D-76128 Karlsruhe, Germany}
\affiliation{${}^4$Low Temperature Laboratory, Aalto University, P.O.~Box 13500, FI-00076 AALTO, Finland.}
\begin{abstract}
    We suggest a scheme to implement a universal set of non-Abelian geometric transformations for a single logical qubit composed of three superconducting transmon qubits coupled to a single cavity. The scheme utilizes an adiabatic evolution in a rotating frame induced by the effective tripod Hamiltonian which is achieved by longitudinal driving of the transmons. The proposal is experimentally feasible with the current state of the art and could serve as a first proof of principle for geometric quantum computing.
\end{abstract}
 \maketitle

\section{INTRODUCTION\label{intro}}
Superconducting qubits, also known as artificial atoms, have emerged
as a promising candidate to achieve quantum computing.\cite{SQRMP}
The properties of these nanoscale systems can be designed to a large
extent, and systems have been found were many logical operations can be
performed within the decoherence time.\cite{koch07,fink09} Superconducting
qubits can be coupled via thin-film microwave cavities~\cite{Blais2004,Wallraff2004} to
allow for two-qubit gates and ultimately for universal quantum
computing. Furthermore, they have the intrinsic scalability of condensed matter
systems and the high-precision measurement features of quantum optical
systems.

Error correction theory predicts that fault tolerant quantum
computing requires of the order of $10^4$ quantum operations with
only a single error on average.\cite{Gottesmann,Knill} Contrary to classical
computation, where gates and errors are discrete, in quantum
computation many small errors can accumulate to an eventual bit or
phase flip. Therefore, enormous accuracy for single gates is
required. Typically, the control parameters cannot be controlled to such
precision and especially the exact timing of control pulses remains
challenging. As a possible solution, holonomic quantum computing was proposed.\cite{zanardi99} 
In this case, the unitary transformations depend only on the path which the
control parameters trace in parameter space, but not on their
timing. Furthermore, random rapidly fluctuating deviations of the actual path
to the desired one cancel to the first order.\cite{dechiara03,
solinas04} Thus, the precision of holonomic quantum gates can possibly be
considerably higher that the precision of the control parameters
itself.

Abelian holonomies, referred to as geometric phases or Berry phases, have been observed in
a wide variety of systems including superconducting
qubits.\cite{leek07, mottonen08} The situation is quite different
for non-Abelian holonomies necessary for universal geometric quantum
computing. Despite several theoretical
proposals~\cite{Choi,Faoro,Brosco,Pirk,solinasPRA10,majorana}, no such adiabatic
transformation has been experimentally observed in superconducting
qubits, nor in any other systems. Here, we present a scheme for the implementation of a non-Abelian holonomy
which is feasible with the devices and methods used in current
experiments on transmon qubits.\cite{fink09}

Our method is based on the much studied tripod Hamiltonian
%$H= \sum_{j=1}^{3} (\Omega_i |0\rangle \langle i |+h.c.)$ ($\hbar =1$)
\begin{equation}
 \hat{H}=  \hbar \sum_{j=1}^{3} (\Omega_i |0\rangle \langle i |+{\rm h.c.}) \hat{=}
         \hbar \left( \begin{array}{cccc}
        0 & \Omega_1 & \Omega_2 & \Omega_3 \\
        \Omega_1^* &0&0&0 \\
        \Omega_2^* &0&0&0 \\
        \Omega_3^* &0&0&0 \\
    \end{array}\right)
 \label{tripod}
\end{equation}
where the $\Omega_i(t)$ are the control parameters (usually referred to as
Rabi frequencies) and the matrix representation is given in the
basis $\{ |i\rangle \}$, $i=0, 1, 2, 3$. The first proposal to
observe non-Abelian transformations in trapped ions was based on 
Hamiltonian~(\ref{tripod}), which is sufficient to implement an arbitray $U(2)$-transformations between the states $\ket1$ and $\ket2$ used as the logical qubit.\cite{unanyan99,duan01} 
Similar structures have been recovered in many
quantum systems and similar implementations have been proposed
\cite{fuentes02, recati02, solinas03, zangh05}, but yet without experimental verification.

Based on recent experiments~\cite{fink09} we propose a way to implement
a universal set of single qubit non-Abelian geometric transformations in a system
of three superconducting transmon qubits coupled to the same
cavity. Each transmon is composed of two superconducting islands
connected by two Josephson junctions, thus forming a superconducting loop.\cite{koch07, fink09} The control parameters are the magnetic fluxes through the loops of each transmon, which can be controlled individually allowing us to adiabatically drive the system along a control cycle. With realistic approximations we are able to obtain an effective tripod Hamiltonian in a rotating frame.

The paper is organized as follows. In Sec.~\ref{sec2}, we introduce
the physical setup of the proposed experiment and in
Sec.~\ref{sec3}, we derive the effective tripod Hamiltonian.
Section~\ref{sec4} is devoted to numerical studies. We conclude the work in
Sec.~\ref{sec5}.

\section{PHYSICAL SYSTEM\label{sec2}}
The superconducting qubits considered here are commonly referred to
as transmons.\cite{koch07,Schreier} Their structure is similar to a
charge qubit, but they have a much larger total capacitance
$C_\Sigma$ such that the ratio of the charging energy
$E_C=e^2/(2C_\Sigma)$ over the Josephson energy $E_J$ is much lower
than unity. This results in a small charge dispersion of the energy
eigenstates, which in turn leads to a significantly reduced
sensitivity to charge noise and much longer decoherence times,
typically of the order of a few microseconds. On the downside, they
have a smaller anharmonicity compared with charge qubits.

Our scheme includes three transmons with frequencies $\varepsilon_i/2\pi,
\;i=1,2,3$. The transmons are coupled to a cavity mode with
frequency $\omega/2\pi$. The combined system is described by the Tavis-Cummings Hamiltonian~\cite{koch07,fink09}
\begin{eqnarray}
    \hat{H} &=& \hbar \omega \hat{a}^\dag \hat{a} + \sum_{i=1}^3 \left[ \half  \hbar \varepsilon_i\hat{\sigma}_z^{(i)} +  \hbar g_i \!\left(\hat{\sigma}_+^{(i)}\hat{a}+\hat{\sigma}_-^{(i)}\hat{a}^\dag\right) \right ], \label{fullham}
\end{eqnarray}
where $g_i/2\pi$ is the transmon--cavity coupling frequency,
$\hat{\sigma}^{(i)}$ are the usual Pauli operators for the $i$-th
transmon, and $\hat{a}$ and $\hat{a}^\dagger$ are the bosonic
annihilation and creation operators for the cavity mode. To arrive
at the above Hamiltonian, we used the rotating wave approximation
(RWA), assuming that the coupling strengths $g_i$ are small compared
with the excitation energies, which will be the case throughout the
paper. 

\begin{figure}[t]
	\includegraphics[width=\linewidth]{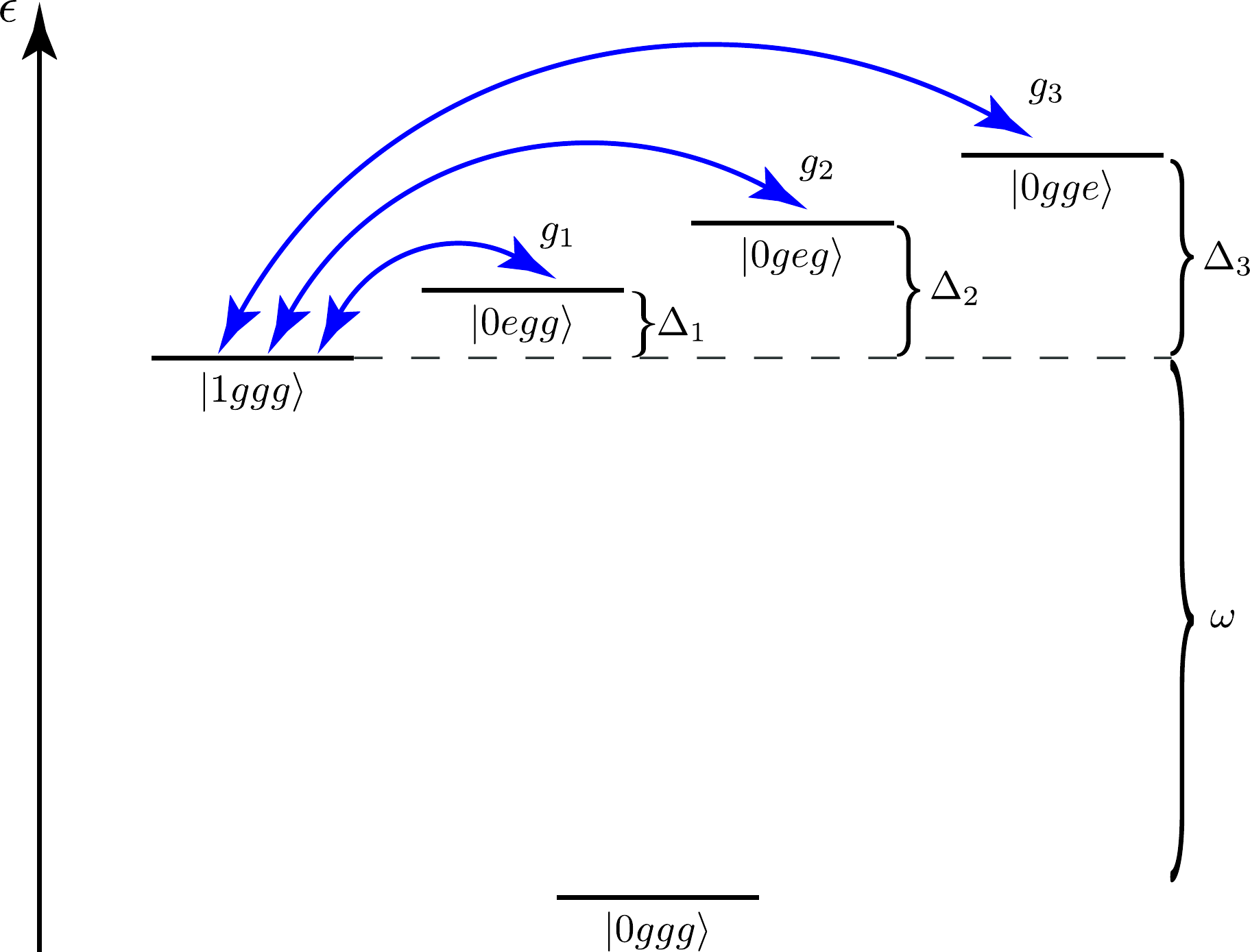}%{new_fig.pdf}
\caption{(Color online) The level structure of the lowest-energy states of the Hamiltonian~\eqref{fullham} are shown. The photon excitation number of the cavity is labeled by $\ket n$, while $\ket g$ and $\ket e$ are the ground and excited states of the three transmons, respectively. The couplings between the one-excitation levels $g_i$ are small compared to the detunings $\Delta_i$. \label{levelstructure}}
\end{figure}

Furthermore, we neglected higher levels of the transmons which is very well
justified, because we do not drive the transmons transversally and therefore do not induce excitations to higher energy levels, and
we initialize the system in the one-excitation subspace.\cite{gs_note}
%%%%
%~\footnote{In fact, after cooling to the ground state, transversal driving is necessary for initialization in the one-excitation subspace. However, as this procedure is very much standard and can be performed with high fidelity it is not discussed further.}.
%
In this case, the only states involved in the dynamics are $\{ |1 ggg
\rangle, |0 egg \rangle, |0 geg \rangle, |0 gge \rangle \}$, where
the first states in the tensor product represents the photon number
in the cavity mode and the states $|e\rangle$ and $|g\rangle$ are the
excited and ground states of each transmon. The Hamiltonian
restricted to this subspace, written in matrix form, reads
\begin{eqnarray}H&=&  \hbar \left(
    \begin{array}{cccc}
        0&g_1&g_2&g_3\\
        g_1&\Delta_1&0&0\\
        g_2&0&\Delta_2&0\\
        g_3&0&0&\Delta_3
    \end{array}\right) , \label{oneham}
\end{eqnarray}
where $\Delta_i=\varepsilon_i-\omega$ is the detuning of the $i$-th transmon from the cavity (see Fig.~\ref{levelstructure}).

The qubit frequencies $\varepsilon_i
$, the detuning $\Delta_i$, and the system--cavity coupling
strengths $g_i$ depend on the Josephson energy $E_{J}^{(i)}(\phi_i)$
and thus on the controllable flux $\phi_i$ though the $i$-th qubit.
The Josephson energy can be written as $E_{J}^{(i)}(\phi_i) =
E_{J{\rm max}}^{(i)} \cos \left (\pi  \phi_i \right)$ where
$E_{J{\rm max}}^{(i)}$ is the maximum Josephson energy and $ \phi_i$
is in units of the flux quantum $h/(2e)$. Explicitly, we have
\cite{koch07,fink09,footnote2}
\begin{eqnarray}
	\varepsilon_i (\phi_i) &=& \sqrt{8 E_{C}^{(i)} E_{J{\rm max}}^{(i)} |\cos \left (\pi  \phi_i \right) |} / \hbar,  \nonumber \\
	g_i(\phi_i) &=& k_i  \left [\cos \left (\pi  \phi_i \right) \right]^{\frac{1}{4}},
	\label{eq:epsilon_g}
\end{eqnarray}
where $k_i$ is a constant depending on the system parameters which can be determined experimentally.

By changing the flux $\phi_i(t)$ through the $i$-th transmon, we can
control the coupling strength $g_i(t)$ as well as the detuning $\Delta_i(t)$.
We separate the the dominant constant contributions [denoted with
superscript $(0)$] which defines the properties of the non-driven system from the small time-dependent ones, which are used to drive the system. We employ the
notation
\begin{eqnarray}
    \mbox{Flux:}\qquad\qquad \phi_i(t) &=& \phi^{(0)}_i + \delta\phi_i(t), \nn\\
    \mbox{Detuning:}\quad\;\,\: \Delta_i(t) &=& \Delta^{(0)}_i + \delta\Delta_i(t), \nn\\
    \mbox{Coupling:}\quad\quad\, g_i(t) &=& g^{(0)}_i + \delta g_i(t), \nn\\
    \mbox{Hamiltonian:}\;\,\, H(t) &=& H^{(0)} + \delta H(t). \nn
\end{eqnarray}
We assume that the flux modulation is small compared to the flux quantum, i.e.\ $\delta\phi_i(t)\ll1$, and use a first-order expansion in $\delta\phi_i$ to obtain the time-dependent quantities. Because the cavity frequency is independent of the flux, we have $\delta \Delta_i=\delta \varepsilon_i$.
Together with $g_i(\phi) \propto \sqrt{\varepsilon_i(\phi)}$
% \begin{eqnarray}
%  \delta \Delta_i (\delta \phi_i) &=& -
%    \frac\pi2 \varepsilon_i^{(0)} \tan \left( \pi \phi^{(0)}_i\right)   \delta \phi_i  \nonumber \\
%      \delta  g_i(\delta \phi_i) &=& - \frac{\pi}{4} g_i^{(0)}   \tan \left( \pi \phi^{(0)}_i\right)  \delta \phi_i.
%      \label{eq:deltaDelta_deltag}
% \end{eqnarray}
we obtain a useful relation between the coupling and detuning
variations
\begin{eqnarray}
    \frac{\delta g_i(\delta\phi_i)}{g_i^{(0)}(\phi^{(0)}_i)} &=& \frac{\delta \Delta_i (\delta \phi_i)}{{2}\varepsilon_i^{(0)}(\phi^{(0)}_i)},\label{relation}
\end{eqnarray}
which is valid up to first order in $\delta\phi_i$. Since typically $g_i\ll\varepsilon_i$, the driving via the flux has
a much smaller effect on the transmon-cavity coupling $g_i$ than on
the detuning $\Delta_i$, and therefore results mainly in longitudnal driving. However, the variations in the detuning induce transitions in higher order perturbation theory,
which we refer to as indirect coupling. Whether, the direct or the indirect
coupling gives the leading contribution to the effective tripod Hamiltonian depends on the ratio $\Delta_i^{(0)}/\varepsilon_i^{(0)}$ and will be studied below in detail.

\section{EFFECTIVE TRIPOD HAMILTONIAN\label{sec3}}

In this section, we show how to obtain an effective tripod Hamiltonian from the Hamiltonian \eqref{oneham}. In particular, the normal tripod approach which solely utilizes the driving of the off-diagonals of the Hamiltonian will not work for our situation, because our control over $g_i$ is rather limited. Nevertheless, we will find that driving the diagonals results in an indirect coupling of the different eigenstates of $H^{(0)}$ which is of the desired tripod form.

To this end, we assume that the
time dependent fluxes $\delta \phi_i (t)$ oscillate with the
frequencies $\omega_i/2\pi$ and we write %Eq.~(\ref{eq:deltaDelta_deltag}) as
\begin{eqnarray}
    \delta\phi_i(t) &=& F_i(t)\cos[\omega_i t + \varphi_i(t)], \nn\\
    \delta\Delta_i(t) &=& L_i(t)\cos[\omega_i t + \varphi_i(t)], \nn\\
    \delta g_i(t) &=& T_i(t)\cos[\omega_it+ \varphi_i(t)],
    \label{eq:L_T_definition}
\end{eqnarray}
where the adiabatically changing amplitudes $L_i(t)$ and $T_i(t)$ are related with the externally controllable magnitude of the flux oscillations $F_i(t)$ by \eqref{eq:epsilon_g}. To realize a universal set of single qubit transformations, also the relative phases $\varphi_i(t)$ of the oscillations need to change adiabatically in time.\cite{duan01}

Anticipating that $\delta H(t)$ drives transitions
between the eigenstates of $H^{(0)}$, we diagonalize $H^{(0)}$. Up
to the first order in $g_i^{(0)}/\Delta_i^{(0)}$, the eigenstates of
$H^{(0)}$ in the basis $\{ |1 ggg \rangle, |0 egg \rangle, |0 geg
\rangle, |0 gge \rangle \}$ are given by
\begin{widetext}
    \begin{eqnarray}
         v_0\approx \left(\begin{array}{c}
            1\\-g_1^{(0)}/\Delta_1^{(0)}\\-g_2^{(0)}/\Delta_2^{(0)}\\-g_3^{(0)}/\Delta_3^{(0)}
        \end{array}\right),\quad
         v_1\approx \left(\begin{array}{c}
            g_1^{(0)}/\Delta_1^{(0)}\\1\\0\\0
        \end{array}\right),\quad
         v_2\approx \left(\begin{array}{c}
            g_2^{(0)}/\Delta_2^{(0)}\\0\\1\\0
        \end{array}\right),\quad
         v_3\approx \left(\begin{array}{c}
            g_3^{(0)}/\Delta_3^{(0)}\\0\\0\\1
        \end{array}\right).\quad 
        \label{eq:v_basis}
    \end{eqnarray}
    The Hamiltonian $H(t)=H^{(0)}+\delta H(t)$ in this  basis assumes the form
    \begin{eqnarray}
        H^D(t)&=& \hbar \left(\begin{array}{cccc}
            0 & \delta g_1(t)-\frac{g_1^{(0)}}{\Delta_1^{(0)}}\delta\Delta_1(t)\; & \delta g_2(t)-\frac{g_2^{(0)}}{\Delta_2^{(0)}}\delta\Delta_2(t)\; & \delta g_3(t)-\frac{g_3^{(0)}}{\Delta_3^{(0)}}\delta\Delta_3(t) \\
                \delta g_1(t)-\frac{g_1^{(0)}}{\Delta_1^{(0)}}\delta\Delta_1(t) \;&E_1+\delta\Delta_1(t)&0&0\\
            \delta g_2(t)-\frac{g_2^{(0)}}{\Delta_2^{(0)}}\delta\Delta_2(t) \;&0&E_2+\delta\Delta_2(t)&0\\
            \delta g_3(t)-\frac{g_3^{(0)}}{\Delta_3^{(0)}}\delta\Delta_3(t) \;&0&0&E_3+\delta\Delta_3(t)
        \end{array}\right)\nn
    \end{eqnarray}
    where the frequencies $ E_i$ can be obtained by perturbation theory
    \begin{eqnarray}
        E_i &\approx& \Delta_i^{(0)}+\frac{2(g_i^{(0)})^2}{\Delta_i^{(0)}}+\sum_{j\neq i}\frac{(g_j^{(0)})^2}{\Delta_j^{(0)}}. \label{star}
    \end{eqnarray}
Here, it is clear that $\delta g_i(t)$ and $\delta\Delta_i(t)$ have to oscillate with frequency $\omega_i=E_i $ to induce an effective coupling. Moving into the rotating frame with respect to the diagonal dominant contribution $H^{\rm diag} = {\rm diag} \{ 0, E_1, E_2, E_3\}$ and using Eq. (\ref{eq:L_T_definition}) we
    obtain
%     \begin{eqnarray}
%         H^D_{km}(t)&=& \hbar\times\left\{\begin{array}{cl}
% \vspace*{4pt}\left( \frac{T_m(t)}2-\frac{g_m^{(0)}L_m(t)}{2\Delta_m^{(0)}}\right)\!\!\left(e^{i \varphi_m}+e^{-2i\omega_m t-i \varphi_m}\right), & \textrm{for } k=0,\;1\leq m\leq 3 \\
% \vspace*{4pt}\left( \frac{T_k(t)}2-\frac{g_k^{(0)}L_k(t)}{2\Delta_k^{(0)}}\right)\!\!\left(e^{-i \varphi_k}+e^{2i\omega_k t+i \varphi_k}\right), & \textrm{for } 1\leq k\leq 3,\;m=0 \\
% \vspace*{4pt} L_{k}(t) \cos[\omega_{k} t+\varphi_{k}],&
% \textrm{for } 1\leq k=m\leq 3 \\
% 0, & \textrm{otherwise}
%         \end{array}\right.\label{poiu}
%     \end{eqnarray}
% 
% 	\begin{equation}
% 		H^{D}_{km}(t) = \left\{
% 		\begin{array}{lcl}
% 			\Omega_{m}(t) \left\{ 1 + e^{- 2 i [\omega_{m} t + \varphi_{m}(t)]} \right\} &\,, & k=0, 1\leq m \leq 3 \\
% 			\Omega_{k}^{*}(t) \left\{ 1 + e^{2 i [\omega_{k} t + \varphi_{k}(t)]} \right\} &\,, & 1 \leq k \leq 0, m = 0 \\
% 			L_{k} \cos{[ \omega_{k} t + \varphi_{k}(t) ]} &\,, & 1 \leq k = m \leq 3 \\
% 			0 &\,, & \mbox{otherwise} 
% 		\end{array} \right. 
% 		\label{eq:5:HOneRotate}
% 	\end{equation}

   \begin{eqnarray}
       H^D(t)&=& \hbar \left(\begin{array}{cccc}
           0 & \Omega_{1} \!\left\{ 1 + e^{- 2 i [\omega_{1} t + \varphi_{1}]} \right\} & \Omega_{2} \!\left\{ 1 + e^{- 2 i [\omega_{2} t + \varphi_{2}]} \right\} & \Omega_{3} \!\left\{ 1 + e^{- 2 i [\omega_{3} t + \varphi_{3}]} \right\} \vspace{2mm}\\
               \Omega_{1}^* \!\left\{ 1 + e^{ 2 i [\omega_{1} t + \varphi_{1}]} \right\} & L_{1} \cos{[ \omega_{1} t + \varphi_{1} ]} &0&0 \vspace{2mm}\\
           \Omega_{2}^* \!\left\{ 1 + e^{ 2 i [\omega_{2} t + \varphi_{2}]} \right\} &0& L_{2} \cos{[ \omega_{2} t + \varphi_{2} ]} &0 \vspace{2mm}\\
           \Omega_{3}^* \!\left\{ 1 + e^{ 2 i [\omega_{3} t + \varphi_{3}]} \right\} &0&0& L_{3} \cos{[ \omega_{3} t + \varphi_{3} ]}
       \end{array}\right). \label{poiu}
    \end{eqnarray}
\end{widetext}
Here, we defined the effective Rabi frequencies
\begin{eqnarray}
    \Omega_i(t)&=&\left(\frac{T_i(t)}2-\frac{g_i^{(0)}L_i(t)}{2\Delta_i^{(0)}}\right)e^{i \varphi_i} \nn\\
    &=& L_i(t)\left( \frac{g_i^{(0)}}{4 \varepsilon_i^{(0)}} - \frac{g_i^{(0)}}{2\Delta_i^{(0)}}\right)e^{i \varphi_i(t)} , \label{effe}
\end{eqnarray}
where \eqref{relation} was used in the second line. In the RWA we can drop the oscillating entries of \eqref{poiu} and we arrive at the desired tripod Hamiltonian~\eqref{tripod}. 

For negative detunings, i.e.\ $\omega>\varepsilon_i^{(0)}$, the direct coupling due to $T_i$ and the indirect coupling due to $L_i$ add up increasing
the strength of the effective coupling. Depending on the ratio
between the detuning and the energy gap we have two different
regimes. If $|\Delta_i^{(0)}| \ll \varepsilon_i^{(0)}$ we are in the
small detuning regime and the second contribution dominates. If
$|\Delta_i^{(0)}| \gg \varepsilon_i^{(0)}$ we are in the large
detuning regime and the first contribution dominates.
Theoretically, both regimes yield the tripod form of the effective
Hamiltonian. Since the different regimes have different requirements
on the experimental setup, which are readily available for small detuning regime, we study this in more detail in below.

We used two approximations in the derivation of the tripod
Hamiltonian. Firstly, $g_i^{(0)}\ll\Delta_i^{(0)}$ was needed to
derive \eqref{poiu}. Although there exist higher order
terms which might seem to destroy the ideal tripod structure, these can all be removed within the RWA. Nevertheless, there are higher order terms resulting in an effective coupling slightly lower than suggested by \eqref{effe} and an optimal driving frequency marginally different from \eqref{star}. Secondly, the RWA requires
$L_i\ll\omega_i\approx\Delta_i^{(0)}$. Both relations limit the effective coupling
strength of the indirect coupling, while the direct coupling is limited by $g_i^{(0)}\ll\varepsilon_i^{(0)}$ which was used to write down \eqref{fullham}. To demonstrate a holonomy with current experimental
limitations (decoherence times, transmon-cavity couplings, flux
driving), one needs to reduce the detuning $\Delta_i^{(0)}$ to the
edge of the validity of the above approximations. This is studied in
the following section.

\section{RESULTS\label{sec4}}

In this section, we verify the validity of above analytical results
with numerical studies. As an example, we choose to work with a
particular implementation of non-Abelian operator proposed in
Refs.~\onlinecite{unanyan99, duan01} in which the Rabi frequencies
$\Omega_i$ are real and parameterized as
\begin{eqnarray}
    \Omega_1&=&\Omega\sin\beta\cos\alpha,\nn\\
    \Omega_2&=&\Omega\sin\beta\sin\alpha,\nn\\
    \Omega_3&=&\Omega\cos\beta.
    \label{eq:Rabi_frequencies}
\end{eqnarray}
Accordingly, in the driving fields in
Eq.~(\ref{eq:L_T_definition}) we take $\varphi_i=0$. We assume $\Omega$ to be constant, while the angles $\alpha$, and $\beta$ change adiabatically in time.

Let us discuss some general properties of the tripod Hamiltonian in
Eq.~(\ref{tripod}). It has two non-degenerate eigenstates, usually
referred to as the bright states, with energies
$\hbar\Omega=\pm\hbar\sqrt{\Omega_1^2+\Omega_2^2+\Omega_3^2}$ and, more
importantly, a degenerate zero-energy subspace
$\mathcal{E}(t)=\mbox{span}\{\ket{D_1(t)},\ket{D_2(t)}\}$. These
so-called dark states are $\ket{D_1(t)}{=} \cos\beta(\cos\alpha\ket1+\sin\alpha\ket2)-\sin\beta\ket3$ and $\ket{D_2(t)} {=} \cos\alpha\ket2-\sin\alpha\ket1$. The
system state $\ket{\psi(0)}$ is initially prepared to be in this
subspace, and if the control parameters are changed adiabatically,
then the system will stay in this subspace during the evolution. In
particular, for a cyclic Hamiltonian $H(0)=H(T)$ the system will
return to the initial subspace $\mathcal E(0)$. However, within this
subspace the state will undergo a non-trivial $U(2)$ transformation which
is the holonomic operator.\cite{Wilczek,Anandan}

\begin{figure}
    \includegraphics[width=8cm]{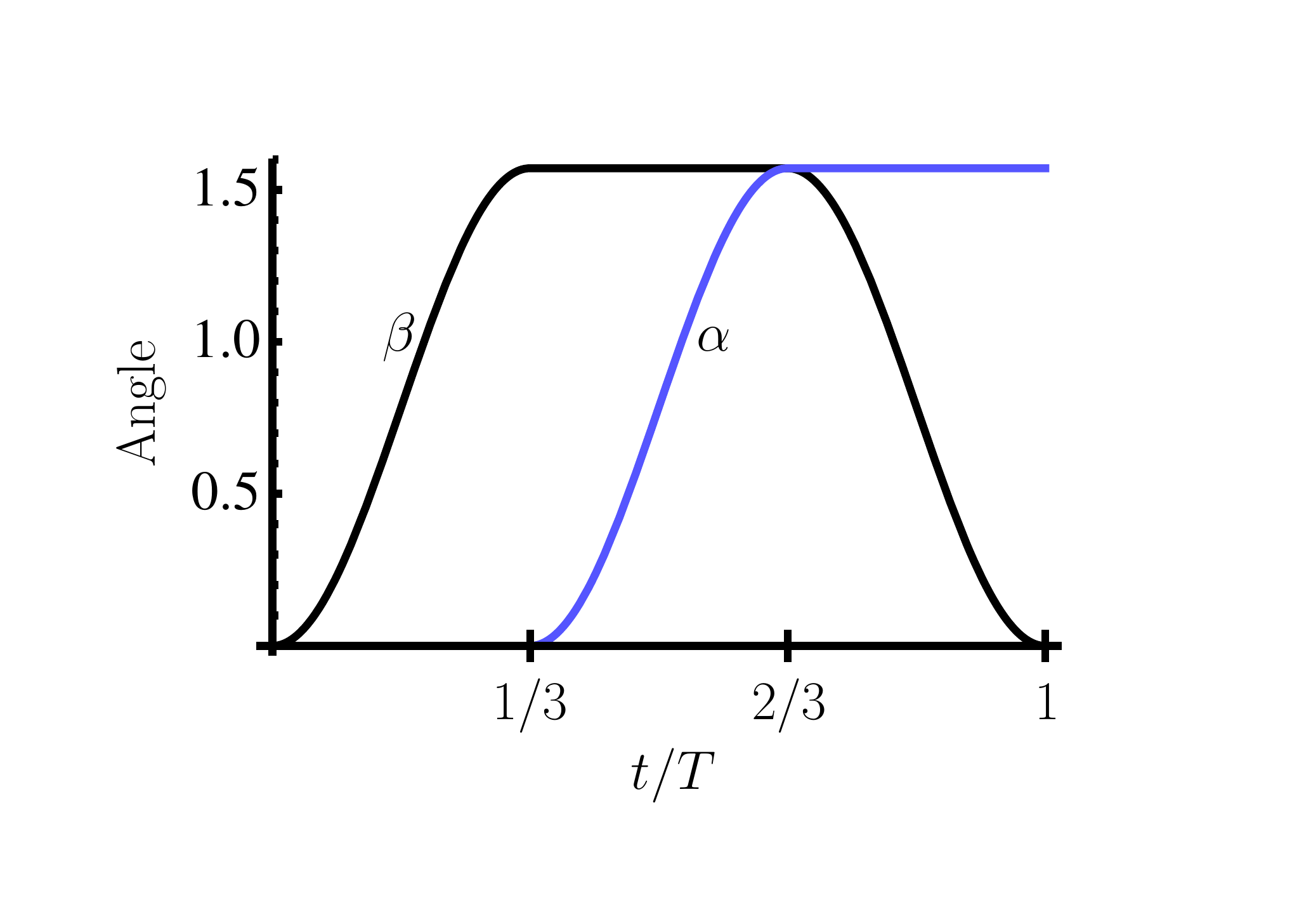}\vspace{-8mm}
    \caption{(Color online) The parameters $\beta$ (black) and $\alpha$ (blue) as a function of time during the adiabatic control cycle.
  \label{figone}}
\end{figure}

The evolution in the parameter space begins and ends at the point
$(\Omega_1, \Omega_2, \Omega_3) = (0, 0, \Omega)$ and by writing
explicitly the dark states using Eq.~(\ref{eq:Rabi_frequencies}), we
obtain that the initial zero-energy subspace is spanned by
$\{|1\rangle, |2\rangle \}$. These states are used as basis states of a logical qubit. An adiabatic change of the angles
according to
\begin{equation}
    (\alpha(t),\beta(t)) : (0,0)\to(0,\pi/2)\to(\pi/2,\pi/2)\to(\pi/2,0), \label{holo}
\end{equation}
results up to a phase factor in a holonomic NOT gate
$U_{\rm hol}=\ket{1}\bra{2}-\ket{2}\bra{1}$ for the logical qubit. Note that because of the spherical
parameterization, the Hamiltonian is cyclic. For better
adiabaticity, we change the angles smoothly using sine functions and constants as shown in
Fig.~\ref{figone}.

To implement this loop in our setup, one can control the
flux driving amplitudes $F_i(t)$ and hence the longitudinal driving amplitude $L_i(t)$ which is directly related to the Rabi frequencies by Eq.~(\ref{effe}). The reference basis is given by
Eq.~(\ref{eq:v_basis}) and the initial degenerate subspace
corresponds to $\ket1\hat=\,v_1$ and $\ket2\hat=\,v_2$. In particular, we assume an initial
state $\ket1$ and, for an ideal transformation, the final state is simply $|\psi_{\rm ideal}(T)\rangle= \ket2$.
Thus, the fidelity, defined as $F(t)= |\langle \psi_{\rm ideal}(t)| \psi (t) \rangle |^2$, after the gate time $T$ is the population of $\ket2$.
For clarity, we only show the populations of $\ket1$ and $\ket2$ in the figures below.

\begin{figure}[t]
	\includegraphics[width=\linewidth]{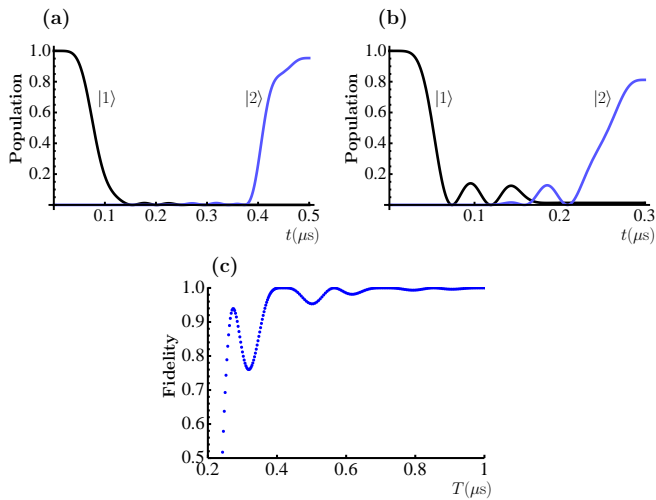}
    \caption{(Color online) The population of the states $\ket1$ (black) and $\ket2$ (blue), obtained from the numerical integration of the Schr\"odinger equation using the effective Hamiltonian in \eqref{tripod} with $\Omega/2\pi=10.5$~MHz. (a)~The gate time $T=0.5$~$\mu$s shows good population transfer, while (b) for $T=0.3$~$\mu$s the smaller fidelity indicates the invalidity of the adiabatic approximation. Panel (c) shows the gate fidelity as a function of the gate time $T$.
        \label{figtwo}}
\end{figure}\vspace{10mm}

We integrate the dynamics of the system using the
ideal tripod Hamiltonian in \eqref{tripod} for different gate times $T$, as shown in Fig.~\ref{figtwo}(a) and (b). The fidelity is plotted over the gate time $T$ in Fig.~\ref{figtwo}(c) and shows the expected approach to unity in the adiabatic limit. The slightly oscillatory behavior observed in Fig.~\ref{figtwo}(c) is typical for adiabatic gates.\cite{florio} However, one should not rely on local maxima of this curve, as their position depends on the precise value of several experimental parameters. Instead, one should use gate times long enough such that even a local minimum provides a sufficiently good fidelity.

\begin{figure}[h]
	\includegraphics[width=\linewidth]{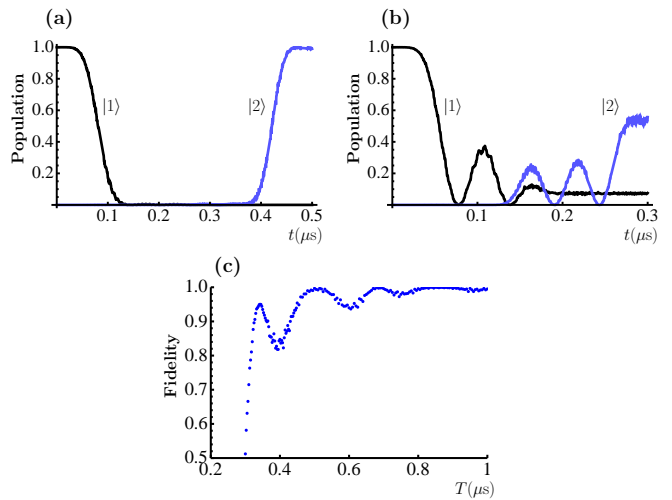}
    \caption{(Color online) Panels (a) and (b): The population of the states $\ket1$ (black) and $\ket2$ (blue), obtained from the numerical integration of the Schr\"odinger equation using the exact Hamiltonian \eqref{oneham}. The parameters are as follows: $\omega/2\pi=5$ GHz, $g_1^{(0)}/2\pi=60$ MHz, $g_2^{(0)}/2\pi=-80$ MHz, $g_3^{(0)}/2\pi=100$ MHz, $\Delta_1^{(0)}/2\pi=-300$ MHz, $\Delta_2^{(0)}/2\pi=-400$ MHz, $\Delta_3^{(0)}/2\pi=-500$ MHz. The longitudinal driving has a restriction of ${\rm max} (L_i)=100$ MHz, resulting in an effective coupling $\Omega/2\pi=10.5$~MHz. The gate times are $T=0.5$~$\mu$s in panel~(a) and $T=0.3$~$\mu$s in panel~(b). Panel (c) shows the fidelity as a function of the gate time $T$.
        \label{figthree}}
\end{figure}

In Fig.~\ref{figthree}, we integrate the exact Hamiltonian of
\eqref{oneham}. By comparing the results with Fig.~\ref{figtwo}, we
can judge whether approximations such as the rotating wave
approximation are satisfied. We
use parameters which yield an effective coupling of
$\Omega/2\pi=10.5$~MHz (10~MHz indirect coupling and 0.5~MHz direct coupling). The results follow closely to the ones obtained
by the effective tripod Hamiltonian, i.e., a gate time of 0.5~$\mu$s
results in a fidelity of almost unity, whereas a gate time of
0.3~$\mu$s is not enough to justify the adiabatic approximation. The fidelity is plotted as a function of the gate time in Fig.~\ref{figthree}(c), which shows much resemblance to the corresponding Fig.~\ref{figtwo}(c) except for a slight rescaling of the gate time. The reason for this rescaling is that $g_i^{(0)}/\Delta_i^{(0)}=0.2$ is not small enough to perfectly justify the approximation $g_i^{(0)}\ll\Delta_i^{(0)}$, and therefore the formula \eqref{effe} slightly overestimates the effective coupling as described in the previous section. 

\begin{figure}[h]
	\includegraphics[width=\linewidth]{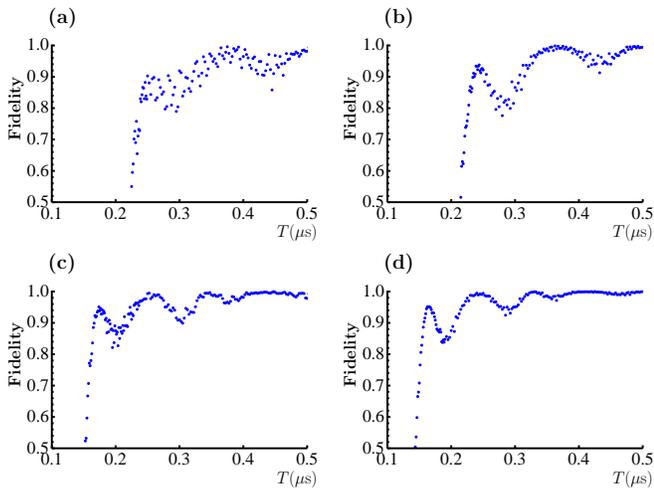}
    \caption{(Color online) Gate fidelity as a function of gate time~$T$. Parameters are choses as in Fig.~\ref{figthree}, but with stronger effective coupling achieved by using half the detuning $\Delta_i^{(0)}$~(a), double the transmon-cavity coupling $g_i^{(0)}$~(b), double the driving $L_i$~(c), and double of all of them~(d). This results in the effective couplings $\Omega/2\pi= 20.5~(a),~ 21~(b),~ 21~(c),~ 22~(d)$~MHz.
        \label{figfour}}
\end{figure}

Because the decoherence time of transmons to date is of the order of a microsecond, one would wish to increase the effective
coupling to achieve faster holonomies. This can be done in various ways as shown in Fig.~\ref{figfour}(a)--(d), always such that \eqref{effe} suggests roughly double the effective coupling compared with Fig.~\ref{figthree}(c). One would expect good fidelities in half the gate time compared with Fig.~\ref{figthree}(c). The easiest and readily available way is to decrease the detunings $\Delta_i$ as shown in Fig.~\ref{figfour}(a). However, the gate fidelity is by
no means as good as expected from the effective coupling. The reason is that the conditions $L_i\ll\Delta_i^{(0)}$ and $g_i\ll\Delta_i^{(0)}$ used in the derivation of the effective Hamiltonian tend to get violated for
decreasing $\Delta_i^{(0)}$. The situation is slightly better in
Fig.~\ref{figfour}(b) and (c), where the higher effective coupling
is achieved by increasing the transmon--cavity coupling $g_i^{(0)}$ and the
driving amplitude $L_i$, respectively. The only way to increase the
effective coupling without affecting the validity of the approximations
is to simultaneously increase $g_i^{(0)}$, $L_i$, and $\Delta_i^{(0)}$, which is
shown in Fig.~\ref{figfour}(d). However, it may be hard to achieve
such high transmon--cavity couplings $g_i^{(0)}$ and driving amplitudes
$L_i$ with current setups.

We would like to add a note on experimental feasibility. The parameters used in Fig.~\ref{figthree} are realistic in an existing experimental setup~\cite{fink09,private} and can be achieved [see \eqref{eq:epsilon_g}], for example, by using transmons with charging energy $E_C=2\pi\hbar\times 280$~MHz, Josephson energy $E_{J\rm{max}}=2\pi\hbar\times 224$~GHz, fluxes $\phi_1^{(0)}=0.48426$, $\phi_2^{(0)}=0.48489$, and $\phi_3^{(0)}=0.48550$, and variations of the fluxes of up to max$(\delta\phi_i)=6.3\times10^{-4}$. To verify the geometric transformation, one has to be able to read out the final state of the system. For this purpose, one increases the detunings considerably such that the energy eigenstates $v_i$ are approximate product states of the cavity and the three transmons. Then, it is sufficient to perform state tomography of the first and second transmon because the holonomy is a transformation between $v_1$ and $v_2$ only. State tomography has been demonstrated for up to three transmons in Refs.~\onlinecite{filipp,dicarlo}.

\section{Conclusion\label{sec5}}
We proposed an experimental scheme for geometric non-Abelian single-qubit gates with superconducting qubits, which could serve as a first step towards geometric quantum computing. Although we did not explicitely take into account the decoherence in our studies, the gate time is within the decoherence time for current experimental setups allowing for proof of principle experiments. The detailed effects of decoherence can be studied along the lines of Refs.~\onlinecite{105,82,82b} and will be presented in a future publication. We note that there could be considerable technical improvements in the near future concerning the decoherence time as well as the driving strength, leading to the possibility to carry out extensive small-scale quantum
computing. We used the NOT gate as an example to calculate the gate
fidelity, but the proposed scheme can be utilized to carry out any
single-qubit transformation~\cite{duan01}.

\acknowledgments{We would like the thank A.~Abdumalikov, S.~Filipp, M.~Pechal, and A.~Wallraff for fruitful discussions, in particular with respect to experimental
parameters. This work was funded under the GEOMDISS project. PS and
MM acknowledge the Academy of Finland and Emil Aaltonen Foundation for
financial support.}


\begin{thebibliography}{99}
\bibitem{gs_note} In fact, after cooling to the ground state, transversal driving is necessary for initialization in the one-excitation subspace. However, as this procedure is very much standard and can be performed with high fidelity it is not discussed further.
\bibitem{footnote2} There is a small correction $-E_C^{(i)}/\hbar$ to the energy splitting $\varepsilon_i(\phi_i)$. This correction done not depend on the flux and can therefore be disregarded, as it merely results in a shift of the detuning $\Delta_i^{(0)}$ which can be countered by adjusting the flux $\phi_i^{(0)}$.
\end{thebibliography}
\end{document}